# Systematic Investigation of Millimeter-Wave Optic Modulation Performance in Thin-Film Lithium Niobate

Yiwen Zhang,[1] Linbo Shao,[2,3] Jingwei Yang,[1] Zhaoxi Chen,[1] Ke Zhang,[1] Kam-Man Shum,[4] Di Zhu,[2,5,6] Chi Hou Chan,[1,4] Marko Lončar,[2] and Cheng Wang[1,4,*]

[1]*Department of Electrical Engineering, City University of Hong Kong, Kowloon, Hong Kong, China*

[2]*John A. Paulson School of Engineering and Applied Sciences, Harvard University, Cambridge, Massachusetts 02138, USA*

[3]*Bradley Department of Electrical and Computer Engineering, Virginia Tech, 1185 Perry Street, Blacksburg, Virginia 24061, USA*

[4]*State Key Laboratory of Terahertz and Millimeter Waves, City University of Hong Kong, Kowloon, Hong Kong, China*

[5]*Institute of Materials Research and Engineering, Agency for Science, Technology and Research (A\*STAR), Singapore, 138634, Singapore*

[6]*zhu_di@imre.a-star.edu.sg*

[*]*cwang257@cityu.edu.hk*

**Abstract:** Millimeter-wave (mmWave) band (30 - 300 GHz) is an emerging spectrum range for wireless communication, short-range radar and sensor applications. mmWave-optic modulators that could efficiently convert mmWave signals into optical domain are crucial components for long-haul transmission of mmWave signals through optical networks. At these ultrahigh frequencies, however, the modulation performances are highly sensitive to the transmission line loss as well as the velocity- and impedance-matching conditions, while precise measurements and modeling of these parameters are often non-trivial. Here we present a systematic investigation of the mmWave-optic modulation performances of thin-film lithium niobate modulators through theoretical modeling, electrical verifications and electro-optic measurements at frequencies up to 325 GHz. Based on our experimentally verified model, we demonstrate thin-film lithium niobate mmWave-optic modulators with a measured 3-dB electro-optic bandwidth of 170 GHz and a 6-dB bandwidth of 295 GHz. The device also shows a low RF half-wave voltage of 7.3 V measured at an ultrahigh modulation frequency of 250 GHz. This work provides a comprehensive guideline for the design and characterization of mmWave-optic modulators and paves the way toward future integrated mmWave photonic systems for beyond-5G communication and radar applications.

## 1. Introduction

Recent years have witnessed a rapid growth of global wireless network traffic. To keep up with the demand of the ever-increasing data capacity, it is an attractive and natural solution to explore new spectral bands that are less congested than currently used microwave bands. The millimeter wave (mmWave) band (i.e., 30 - 300 GHz) is particularly interesting since it has higher frequencies than microwave bands and therefore allows for much larger channel bandwidths according to Shannon's theorem. mmWaves are also important for short-range radar and sensing applications [1]. However, the transmission and processing of mmWave signals is challenging (and costly) due to exacerbating metallic losses as well as the gain-bandwidth trade-off of traditional electronic components at high frequencies. mmWave photonics is a promising solution to address these issues in a cost-effective manner, since it allows generation, transmission and processing of mmWave signals in the low-loss optical domain, in analogy to the concept of microwave photonics [2].

For almost all mmWave-photonic systems, a key component is an electro-optic modulator that could efficiently convert mmWave signals into the optical domain [3, 4]. However, it is a highly non-trivial task for electro-optic modulators to operate at high mmWave frequencies, in particular the range between 100 - 300 GHz. The electro-optic bandwidths of semiconductor-based (e.g., Si and InP) modulators could barely reach this range, limited by their carrier lifetimes [5, 6]. Other emerging material platforms like electro-optic polymer or graphene have shown modulation bandwidths into the terahertz regime [7, 8], yet their long-term stability and scalability remain to be proven. Lithium niobate (LN, $LiNbO_3$) is a promising candidate for mmWave-optic modulation purposes, since the Pockels-based electro-optic effect in LN intrinsically happens on femtosecond timescales and the material itself has been industry proven for decades [9]. However, the bandwidths of traditional off-the-shelf LN modulators are typically limited to < 35 GHz, since the weak electro-optic interactions in ion-diffused waveguides lead to the requirement of long modulation electrodes and large RF losses, especially at high frequencies.

Benefiting from the development of ion-slicing and nanofabrication technologies, the thin-film LN (TFLN) platform has recently shown great promise for pushing the operation bandwidths of LN modulators into the mmWave regime while exhibiting smaller device footprints and lower power consumption [9]. Owing to the much larger refractive index contrast and better confined optical mode in TFLN, the modulation electrodes could be placed much closer to the optical waveguide, leading to substantially increased electro-optic overlap and shorter modulation electrodes required. As a result, a number of high-performance TFLN modulators have been developed, demonstrating low half-wave voltages ($V_\pi$) [10, 11], high modulation bandwidths around 100 GHz [12-17], as well as ultrahigh linearity [18]. The high-performance TFLN modulators could potentially be further integrated with frequency comb sources [19, 20], tunable filters [21, 22] and low-loss delay lines [23] on the same platform for future microwave- and mmWave-photonic applications. While many of these demonstrated modulators have theoretically predicted modulation capabilities much beyond 100 GHz [14-16], the experimental demonstrations are limited. In particular, it remains unclear whether the performances would match theoretical predictions due to the lack of electrical and electro-optic measurements at ultrahigh frequencies. In this range, the electro-optic efficiency is very sensitive to velocity-matching condition and RF losses, and any deviation from the theoretical models can negatively impact the modulator performance. For example, Mercante et al. have measured the electro-optic responses of TFLN modulators at frequencies up to 500 GHz showing good potential, but the measured electro-optic response substantially differs from the theoretically predicted curve since electrical measurements (and in turn, model verifications) are limited to 110 GHz [15]. The measured 3-dB electro-optic bandwidth of the device is also limited to ~ 40 GHz. The lack of a comprehensive design guideline with electrical and electro-optic measurements-based verifications has become a major hurdle for TFLN modulators toward future mmWave-photonic applications.

This work provides a systematic investigation of the velocity- and impedance-matching conditions and RF losses of TFLN modulators based on electrical and electro-optic measurements at frequencies up to 325 GHz. Our results show excellent agreement not only between the simulated and measured electrical parameters (i.e., RF index, impedance and loss), but also between the electro-optic responses predicted from measured electrical parameters and the actually measured electro-optic responses. Based on the proposed design guidelines, we experimentally demonstrate a 5.8 mm long Mach-Zehnder modulator on TFLN with 3-dB and 6-dB electro-optic bandwidths of 175 GHz and 295 GHz, respectively (with reference to 1 GHz). The measured RF $V_\pi$ is as low as 7.3 V at an ultrahigh frequency of 250 GHz nearly at the upper bound of the mmWave spectrum, showing practical relevance for the majority of mmWave photonics applications.

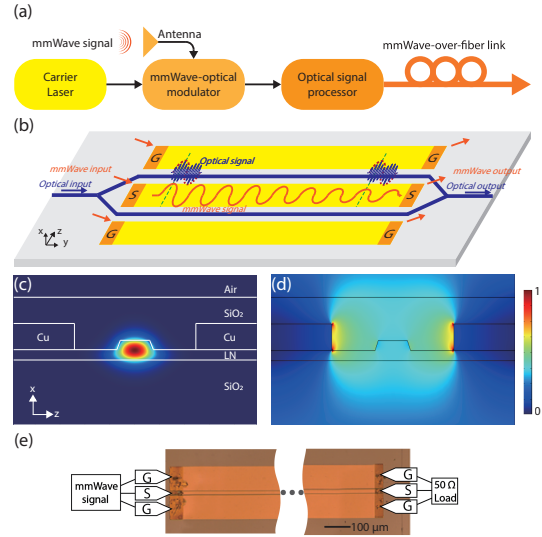

Fig. 1. (a) Schematic illustration of a future mmWave-photonic system, at the heart of which sits the mmWave-optic modulator that converts mmWave signals into the optical domain. (b) Schematic of the TFLN mmWave-optic modulator, where velocity matching between the optical and mmWave signals, impedance matching and RF loss conditions determine the ultimately achievable modulation bandwidths. (c) Simulated optical mode profile ($E_z$) in the TFLN rib waveguide. (d) Simulated mmWave profile ($E_z$) at a frequency of 300 GHz. (e) Optical micrograph of a fabricated device (the darker regions at the two ends are exposed areas for electrical contacts, whereas other parts of the chip are cladded with silicon dioxide).

## 2. mmWave-optic modulator design methodology

Figure 1(a) illustrates an envisioned a future mmWave-photonic system, where mmWave signals received by an antenna could be efficiently converted into optical domian via a modulator, filtered and processed using optical techniques, and/or transmitted over long distances through optical fiber links. An electro-optic modulator with a broad bandwidth covering the entire mmWave range is vital of such a system. Importantly, the peripheral components, including antennas [3], lasers [24], optical filters [21, 22], amplifiers [24, 25] and delay lines [23], have all been demonstrated on the LN platform and could potentially allow the integration of this entire system on the same chip.

Figure 1(b-c) depicts the general structure of our mmWave-optic modulator, which follows the same design as most broadband TFLN modulators demonstrated to date [26]. The modulator consists of a ground-signal-ground (GSG) copper transmission line

and an optical Mach-Zehnder interferometer (MZI). The modulator is based on an x-cut TFLN on top of thermal oxide on Si substrate, such that a combination of transverse-electric (TE) optical modes and in-plane electrodes allows the exploitation of the largest electro-optic tensor component $r_{33}$ of LN.

There are two main performance metrics for a mmWave-optic modulator, which we focus on optimizing in the rest of this paper: (1) the electro-optic $S_{21}$ (EO $S_{21}$) parameter, which depicts the relative modulation efficiency roll-off with reference to the DC or low-frequency (1 GHz in this work) electro-optic response; and (2) the RF $V_\pi$, which indicates the absolute modulation efficiency at a certain frequency. The 3-dB and 6-dB bandwidths of a modulator refer to the frequencies at which EO $S_{21}$ drops by the respective dB numbers. At 3-dB and 6-dB points, the RF $V_\pi$ values are 1.414× and 2× the low-frequency $V_\pi$ respectively, since the input electrical power scales quadratically with modulation voltage. Therefore both 3-dB and 6-dB bandwidths are relevant metrics for practical applications.

The modulation performance at high RF frequencies is mainly limited by three factors [Fig. 1(b)]. First, the phase velocity of mmWave should be matched with the optical group velocity, such that the same electrical signal could be continuously applied to the propagating optical pulse resulting in an efficient modulation. Second, the mmWave loss should be minimal such that the modulation signal could remain significant throughout the modulator. Third, the impedance of the transmission line should be matched with that of external electronics (typically 50 Ω) to avoid power reflection at the input port. Quantitatively, the effective modulation voltage (normalized to input voltage) averaged along the transmission line can be expressed as:

$$m(\omega) = \left| \frac{2Z_{in}}{Z_{in}+Z_C} \right| \left| \frac{(Z_C+Z_0)F_+ + (Z_C-Z_0)F_-}{(Z_C+Z_0)e^{\gamma_m L}+(Z_C-Z_0)e^{-\gamma_m L}} \right| \quad (1)$$

where $\omega$ is the mmWave frequency, $Z_{in} = Z_0 \frac{Z_C + Z_0 \tanh(\gamma_m L)}{Z_0 + Z_C \tanh(\gamma_m L)}$ is the transmission line input impedance, $Z_0$ is the characteristic impedance of the transmission line, $Z_C$ is the impedance of the source and termination (usually 50 Ω), $F_\pm = (1 - e^{\pm \gamma_m L - j\frac{\omega}{c} n_o L}) / (\pm \gamma_m L - j\frac{\omega}{c} n_o L)$ corresponds to the forward/backward propagating waves, $\gamma_m = \alpha_m + j\frac{\omega}{c} n_m$ is the microwave propagation constant ($n_m$ is the mmWave effective index, $\alpha_m$ is the loss coefficient), $n_o$ is the optical group refractive index, $L$ is the modulation length, $c$ is the speed of light in vacuum.

The relative modulation efficiency, EO $S_{21}$, is simply $|m(\omega)|^2$ in dB scale since it relates the power roll-off, which could be written as:

$$\text{EO } S_{21} = 10\log\left[(1-H)^2 \frac{|S_{21}|^2 - 2\cdot|S_{21}|\cdot\cos(\beta_{opt}^\mu L)+1}{(\ln|S_{21}|)^2 + (\beta_{opt}^\mu L)^2}\right] \quad (2)$$

$$\beta_{opt}^\mu = \frac{\omega}{c}(n_m - n_o) \quad (3)$$

$$H = \frac{Z_C - Z_{in}}{Z_C + Z_{in}} \quad (4)$$

where $S_{21}$ is the electrical forward transmission coefficient in linear scale, which represents the contribution from RF loss; $\beta_{opt}^\mu$ is the wavevector (velocity) mismatch term as defined in Eq. (3); $H$ characterizes the mmWave reflection due to mismatch between the input impedance of the transmission line, $Z_{in}$, and that of the driving circuit, $Z_C$, as defined in Eq. (4).

Ideally, when velocity-matching and impedance-matching conditions are met, both the mismatch terms $H$ and $\beta_{opt}^\mu L$ are equal to zero. The EO $S_{21}$ is then only determined by the electrical $S_{21}$ value. Numerically, the 3-dB and 6-dB electro-optic bandwidths correspond to the frequencies at which the electrical $S_{21}$ responses roll-off by 6.41 dB and 13.8 dB, respectively. The EO $S_{21}$ response rolls off slower than electrical $S_{21}$ since the former captures the average modulation voltage along the transmission line. In contrast, the latter corresponds to the transmitted electrical power at the output port.

The absolute RF $V_\pi$ values could then be calculated from the EO $S_{21}$ parameter and the low-frequency $V_\pi$ value using:

$$V_{\pi,\text{RF}} = V_{\pi,\text{LF}} \times 10^{-\text{EO } S_{21}/20} \quad (5)$$

The DC/low-frequency $V_\pi$ values in most TFLN modulators reported to date follow voltage-length products ($V_\pi L$) in the range of 2-3 V·cm [11, 13, 14], in order to achieve a good balance between electro-optic overlap and metal-induced optical loss. The $V_\pi L$ product in the devices characterized in this work is 2.3 V·cm at DC and 2.67 V·cm at 1 GHz, consistent with other previous demonstrations. The $V_\pi L$ product could potentially be further reduced by material engineering or novel modulation structures, while the design methodologies introduced in this paper for mmWave-band operation shall remain largely the same.

Following the above theory, the requirements and trade-offs below should be considered to design a broadband TFLN mmWave-optic modulator:

(1) Velocity matching between optical and mmWave could be achieved by fine-tuning the buried and cladding oxide thickness. A thicker buried (cladding) oxide layer pushes the mmWave mode [Fig. 1(d)] away from the high-index Si substrate (low-index air), therefore decreasing (increasing) the mmWave effective index.

We numerically simulate the electrical characteristics of the mmWave transmission line using Finite Element Methods (FEM, Ansys HFSS) and extract the mmWave effective index ($n_m$) from the numerically converged propagation constant. Figure 2(a) shows the simulated mmWave dispersion curve for different cladding $SiO_2$ thicknesses. Our model assumes a dielectric constant of 5.5 for the cladding $SiO_2$ from plasma-enhanced chemical vapor deposition (PECVD), which is inferred from the actual electrical measurements to be discussed in Section 3. The optical group index ($n_o$) does not change significantly for the fundamental TE mode in our TFLN waveguides, and is simulated to be 2.26 [dashed green line in Fig. 2(a)] (Ansys Lumerical Mode) and is insensitive to the buried/cladding oxide thickness. At a cladding thickness of 600 nm and a buried oxide thickness of 2 μm [solid yellow line in Fig. 2(a)], the mmWave transmission line exhibits phase velocities well matched with the optical group velocity throughout the mmWave band. This set of parameters is used for the actual devices in this work.

(2) The signal width of the mmWave transmission line should be carefully chosen to balance the trade-off between impedance matching and RF loss. According to common transmission line theory, the characteristic impedance $Z_0 = (L/C)^{1/2}$ is determined by the capacitance per unit length $C$ and the inductance per unit length $L$. Using a wider signal line could effectively reduce the Ohmic loss, but at the same time increases the capacitance between the signal and ground and reduces the inductance, leading to lower and often unmatched impedance.

We use the same HFSS model to extract the RF loss and characteristic impedance at 250 GHz as a function of the signal width, which clearly shows a trade-off between the two [Fig. 2(c)]. Due to meshing-related uncertainties of the simulation software, the RF loss values fluctuate slightly, which does not affect the overall trend. It should be noted that impedance mismatch within a certain range does not significantly affect the modulation performance. For example, a 40-Ω characteristic impedance corresponds to ~ 11% drop in the modulation voltage. As a result, the best electro-optic response may not occur at the exact impedance-matched point. The green line of Fig. 2(c) shows our calculated electro-optic responses at 250 GHz based on the calculated RF loss and impedance (assuming velocity matching), indicating a maximum electro-optic response at a signal width of 20 μm, which we use in the actual devices. More advanced transmission line structures like capacitive loading have recently shown the capability to break this trade-off at frequencies below 100 GHz [14, 16, 27, 28], which could potentially be further extended into the upper mmWave bands using smaller capacitive loading periods to increase the cut-off frequency. Nevertheless, the design methodology and characterization techniques introduced in this paper shall still apply to such more advanced modulator designs.

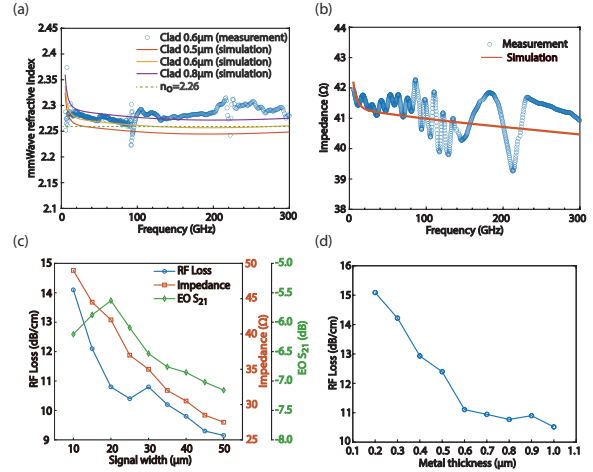

Fig. 2. (a) Simulated mmWave effective (phase) indices for various cladding thicknesses (solid lines) and the actual indices extracted from measured s-parameters (circle), as functions of frequency. The green dashed line shows the simulated optical group index, indicating good velocity matching in the fabricated device. (b) Simulated and extracted characteristic impedance of the device. (c) Simulated RF loss and characteristic impedance, as well as calculated EO $S_{21}$ (assuming velocity matching) for various signal widths at 250 GHz. (d) Simulated RF loss versus metal thickness at 250 GHz.

(3) A thicker metal layer could be used to reduce the Ohmic loss of the transmission line, as far as the fabrication process allows. However, metal thicknesses beyond 800 nm do not quite benefit operations at upper mmWave frequencies since the skin depth at, for example 250 GHz, is only 150 nm, as Fig. 2(d) shows. In our devices to be discussed next, the metal thickness is chosen to be 800 nm. Metal thickness also slightly shifts the mmWave velocity and impedance, which could be fine-tuned using the methods discussed in the above points 1-2.

(4) A longer modulation length leads to faster EO $S_{21}$ roll-off due to larger RF losses, thus smaller 3-dB and 6-dB bandwidths. However, for practical applications at certain mmWave frequencies, the more important parameter is the absolute modulation efficiency, i.e., RF $V_\pi$. In the ideal case where velocity is perfectly matched, a longer modulation length always leads to lower RF $V_\pi$, since even the much-attenuated sections still contribute to the total modulation effect. However, longer electrodes in an actual device are more sensitive to any velocity mismatches, as can be seen in Eq. (2). In this paper, we evaluate modulators with two different lengths, i.e., 5.8 mm and 10.8 mm, which show a good balance between the achievable RF $V_\pi$ and the tolerance to fabrication variations.

We note that actual electrical parameters often deviate from the simulated values due to deviations in actual fabrication parameters and material properties. Therefore, it is important to perform a careful electrical evaluation of the fabricated mmWave transmission lines and fine-tune the theoretical model to achieve the desired modulation performances in the full electro-optic devices, which we discuss next.

## 3. Experiments

A. Device fabrication

The devices are fabricated on a TFLN wafer from NANOLN consisting of a 600 nm thick x-cut LN bonded on top of thermal oxide (2 μm thick) on a 500-μm-thick silicon substrate. We use electron-beam lithography (EBL) to define the optical waveguide patterns in Hydrogen Silsesquioxane (HSQ) and transfer the patterns into TFLN by dry etching 300 nm of the LN film using argon ion ($Ar^+$)-based reactive ion etching (RIE) process [13]. Then, we deposit a metal layer (800-nm-thick copper terminated with 30-nm-thick gold on top) to form the transmission lines through aligned photolithography, evaporation and lift-off processes. The transmission line signal width is 20 μm and the gap between electrodes is 5 μm. The rib optical waveguide is 1.2 μm wide on top in the modulation region and 0.8 μm wide for routing, with a 300 nm thick slab. The chip is cladded with 0.6 μm thick silica by PECVD. The silica cladding is selectively removed at probe contact areas by another photolithography process followed by RIE etching. Finally, the edge of the chip is diced and polished to enhance the coupling between fiber and chip. Figure 1(e) shows the top view of a fabricated mmWave-optic modulator under an optical microscope.

B. Characterizations of device electrical properties.

We first perform a detailed electrical analysis of the fabricated mmWave-optic modulators. We measure the reflection ($S_{11}$) and transmission ($S_{21}$) s-parameters of the transmission line from 10 MHz to 325 GHz using an Agilent vector network analyzer (VNA) and frequency extension modules in the 10 MHz-69 GHz, 65-110 GHz, 90-140 GHz and 220-325 GHz bands, respectively. The VNA is calibrated using short-open-load-thru (SOLT) standards. A pair of mmWave GSG probes is used to launch mmWave signals into the input port of the transmission line and collect them from the output port. Figure 3(a) shows the measured $S_{11}$ and $S_{21}$ parameters of the 5.8-mm and 10.8-mm devices, respectively. As expected, the transmitted power rolls off at higher frequencies, with a measured loss of 1.3 dB/mm at 250 GHz. From the measured s-parameters, we can extract the electrical loss coefficient $\alpha$ of the transmission line, as shown in Fig. 3(b). The electrical loss comes from two origins: the conductor loss $\alpha_c$, which is typically proportional to the square root of frequency, and the dielectric loss $\alpha_d$, which goes linearly with frequency. The total electrical loss for an electrode length $L$ could then be modelled as $\left(\alpha_c\sqrt{f}+\alpha_d f\right)L+A$, where $A$ is the intercept loss at DC originated from impedance mismatch and is equal to 2.6 dB in this case. Based on our measured results, we estimate that $\alpha_c$=0.042 dB cm$^{-1}$ GHz$^{-1/2}$ and $\alpha_d$=0.025 dB cm$^{-1}$ GHz$^{-1}$. The main loss mechanism at frequencies below 50 GHz is conductor loss, consistent with previous reports [14]. However, at upper mmWave bands, the dielectric loss contribution becomes more significant and should not be ignored, as the gap between the two fitted lines in Fig. 3(b) shows. The measured electrical loss is even larger than our model at above 200 GHz, possibly because the dielectric loss tangent increases at higher frequencies while our model assumes a fixed loss coefficient. At these frequencies there also exist resonance-like features likely due to reflections at the two ends of the transmission lines. The extracted $\alpha_d$ gives rise to an effective dielectric loss tangent of 0.025 for this transmission line mode, which is consistent with the loss tangents of the dielectrics involved here, i.e., 0.025 for silicon [29], 0.008 for lithium niobate [30], and 0.000042 for $SiO_2$ [31].

Apart from the RF losses, it is important to extract the actual mmWave velocity and impedance in the fabricated devices, in order to fulfill and verify the design guidelines in Section 2. The mmWave index $n_m$ is extracted by $n_m = \frac{\beta c}{2\pi \omega}$ ($\beta$ is the propagation constant extracted from the measured phase response). The impedance is extracted by $Z_{in} = Z_C \sqrt{\frac{(1+S_{11})^2 - S_{21}^2}{(1-S_{11})^2 - S_{21}^2}}$. The circles in Fig. 2(a-b) shows the extracted $n_m$ and $Z_{in}$ values at frequencies up to 300 GHz, both of which match well with the designed numbers. The difference between the measured and target $n_m$ is 0.02 at 150 GHz and 0.04 at 250 GHz, allowing for a reasonably well-matched velocity within the entire frequency range of interest. The extracted $Z_{in}$ is 41.5 Ω at 250 GHz, which is also close to the simulated result (40.5 Ω). The extracted values see relatively large fluctuations at certain frequencies since the extension modules exhibit degraded performances near the edges of each measurement band. The remaining differences between the measured and simulated $n_m$ and $Z_{in}$ values could result from variations in the fabricated device's structural parameters and measurement uncertainties. Nevertheless, the ability to extract the actual velocity, impedance and loss information of the mmWave transmission line serves as an invaluable asset for predicting and explaining the modulation performances at mmWave frequencies, as we show next.

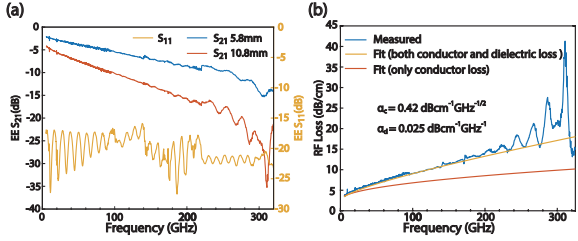

Fig.3. (a) Measured s-parameters of mmWave transmission lines with lengths of 5.8 mm and 10.8 mm. (b) Extracted and fitted electrical loss coefficient of the transmission line as a function of frequency.

C. Characterizations of device electro-optic properties.

We measure the electro-optic responses of our TFLN mmWave-optic modulators using the setup shown in Fig. 4. A wavelength-tunable laser source (Santec TSL-550, 1500-1630 nm) is used to input light in the telecom L-band. A 3-paddle fiber polarization controller (FPC) is used to ensure TE mode excitation. Light is coupled into and out from the chip using a pair of lensed fibers. Measurements are separately conducted in four frequency bands. For the lowest band (< 67 GHz), modulation electrical signals are directly generated from a RF generator (MG3697C, Anritsu). For higher frequencies (> 67 GHz), microwave signals are generated by up-conversion, where 250 kHz – 20 GHz signals (Agilent E8267D) are up-converted and amplified by frequency multipliers in the respective bands, i.e., 65 – 110 GHz, 90 – 140 GHz and 220 – 325 GHz. A pair of mmWave probes are used to deliver the modulation signal to the input port of the transmission line, and to terminate the output port with a 50-Ω load. The modulator is biased at the quadrature point, resulting in an output optical signal with two sidebands separated from the carrier by the mmWave frequency [inset of Fig. 4(a)]. The electro-optic response is tested by monitoring the power ratio between the sideband and the carrier, using an optical spectrum analyzer (OSA), which we define as the normalized sideband power $P_s$:

$$P_s = \frac{P_{sideband}}{P_{carrier}} \quad (6)$$

In Fig. 4(a), we could see that the upper and lower sidebands show slightly different powers. This is because at ultrahigh modulation frequencies, the modulated sidebands are significantly away from the carrier such that they experience a shifted bias point in the unbalanced MZI. In this case, we take the average power of the two sidebands, which allows the below analysis to hold in presence of the unbalanced sidebands.

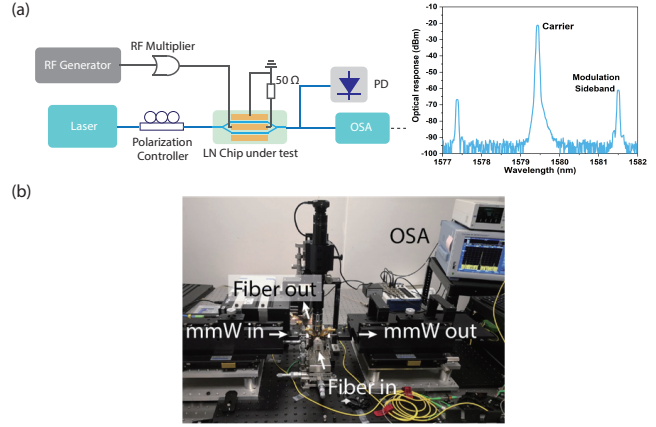

Fig. 4. Schematic diagram (a) and photo (b) of the measurement setup for characterizing electro-optic responses at frequencies up to 325 GHz.

We could therefore calculate the RF $V_\pi$ using Eq. (7) [15]:

$$V_{\pi, RF} = \frac{1}{4}\pi V_p \sqrt{P_s} \quad (7)$$

where $V_p$ is the peak voltage of the input mmWave signal, which is measured using a mmWave power meter at each frequency point and carefully calibrated by excluding the losses from the input mmWave waveguide and the probe.

Based on the extracted RF $V_\pi$ values and the $V_\pi$ at 1 GHz ($V_{\pi, LF}$), we could further calculate the EO $S_{21}$ using Eq. (5).

Figure 5 shows the extracted RF $V_\pi$ (a-b) and EO $S_{21}$ (c-d) at frequencies up to 325 GHz for the 10.8-mm and the 5.8-mm device (blue dots), plotted together with the ones calculated from the measured electrical parameters (red lines) in Section 3B following Eq. (2). The directly measured electro-optic responses show more fluctuations mostly due to uncertainties in the quoted $V_p$ in Eq. (7). The mmWave multiplexer often generates spurious harmonics away from the intended frequency, which may also be counted by the power meter. Nonetheless, the direct electro-optic measurements show excellent agreement with the calculated results, thanks to the precise measurements of velocity, impedance and mmWave loss we have performed.

Our measurement results indicate that the 5.8-mm device features ultrahigh 3-dB and 6-dB electro-optic bandwidths of 170 GHz and 295 GHz, respectively [Fig. 5(d)]. The bandwidths of the 10.8-mm device are relatively lower, i.e.,100 GHz (3 dB) and 175 GHz (6 dB), due to more severe RF attenuation [Fig. 5(c)]. Still, the 10.8-mm device provides lower RF $V_\pi$ than that of the 5.8-mm device throughout the entire mmWave band for the reasons discussed in Section 2 and could be more appealing in most application scenarios. The measured RF $V_\pi$ values are 7.3 V for the 10.8-mm device and 8 V for the 5.8-mm device [Fig. 5(a-b)] at an ultrahigh mmWave frequency of 250 GHz, making these devices highly relevant for practical applications. In comparison, the RF $V_\pi$ of a typical commercial LN modulator could be as high as 15 V at 100 GHz (3.8 V in our device), as measured in [32].

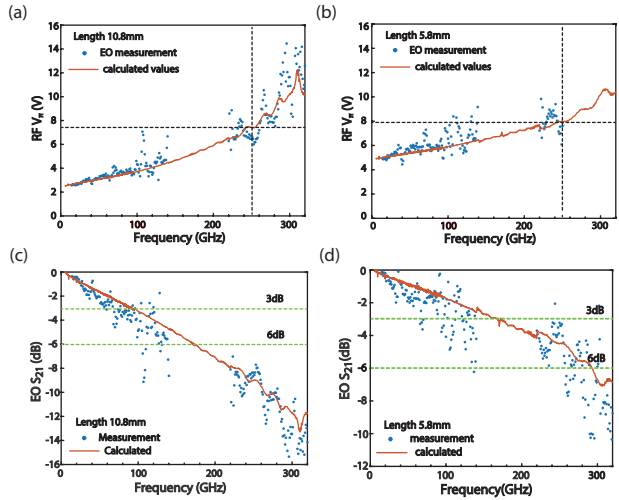

Fig. 5. (a)-(b) Measured and calculated modulator RF $V_\pi$ of the 10.8 mm (a) and 5.8 mm (b) devices. (c)-(d) Calculated and measured electro-optic responses of the 10.8 mm (c) and 5.8 mm (d) devices. The measured lines are extracted directly from the electro-optic sideband measurements. The calculated lines are based on the nm and $Z_{in}$ values from the measured electrical s-parameters.

## 4. Discussions

The absolute RF $V_\pi$ is usually more important for practical applications as it determines the actual mmWave-optic conversion efficiency. As presented above, a longer modulator length should always lead to lower $V_\pi$ as long as the velocity matching condition could be fulfilled, which, however, becomes increasingly more vulnerable for longer electrodes and higher frequencies. Figure 6 shows the calculated RF $V_\pi$ values as functions of device length and velocity (index)/impedance mismatch at 250 GHz. The current devices (yellow dots) operate along the yellow line with a slight index mismatch of $\Delta n = 0.04$. Along this line, the RF $V_\pi$ first decreases and then increases as the modulator length increases, indicating that there exists an optimal modulation length that provides the lowest RF $V_\pi$ (in this case, 6.27 V at 1.15 cm). If perfect velocity matching could be achieved, the RF $V_\pi$ shall monotonically decrease following the red line, potentially achieving an RF $V_\pi$ of 4.17 V at 250 GHz for a device length of 1.5 cm. This performance envelope could be further improved by satisfying the velocity- and impedance-matching conditions simultaneously (blue line), which is currently not possible for a reasonably wide signal line, but could potentially be achieved using a capacitive-loaded electrode structure with small enough periods to extend the cutoff frequencies into upper mmWave frequencies [14, 16, 27, 28].

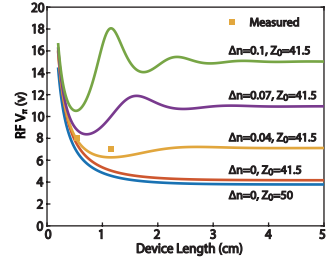

Fig. 6. Simulated RF $V_\pi$ vs. device length for various velocity- and impedance-mismatch conditions at 250 GHz. The yellow dots correspond to the 5.8-mm and 10.8-mm devices in this work.

## 5. Conclusions

In this paper, we report a detailed theoretical and experimental analysis of the mmWave-optic modulation performances of TFLN modulators at frequencies up to 325 GHz. We show that the ability to reliably extract the actual mmWave velocity and impedance in the fabricated devices, is key to an accurate and systematic design optimization at these ultrahigh frequencies. We show that an optimized 5.8 mm long modulator could provide a measured 3-dB electro-optic bandwidth of 170 GHz and a 6-dB bandwidth of 295 GHz. The devices also show RF half-wave voltages as low as 6.3 V at an ultrahigh modulation frequency of 250 GHz. The design and characterization methodologies in this work could be readily applied to more advanced modulator architectures as well as other material platforms. The ultra-broadband low Vπ mmWave-optic modulators could become key elements in future mmWave systems for telecommunication, short-range radar and sensors applications.

**Funding.** National Natural Science Foundation of China (61922092); Research Grants Council, University Grants Committee (CityU 11204820, CityU 21208219, T42-103/16-N); Croucher Foundation (9509005); City University of Hong Kong (9610402, 9610455); Central Research Fund (CRF), Agency for Science, Technology and Research (A*STAR); Harvard Quantum Initiative (HQI) Postdoctoral Fellowship

**Disclosures.** The authors declare no conflicts of interest.

**Data availability.** The data in this study are available from the corresponding authors on reasonable request in the paper.